\documentclass[apjl]{emulateapj}
\usepackage{apjfonts}
\newcommand{\kms}{{\,\rm km\,s}^{-1}}

\renewcommand{\mag}{\mbox{$\;$mag}}
\begin{document}

\title{THE SUPERNOVA I\lowercase{a} 2011\lowercase{fe} IN M101, ITS
  TIP OF THE RED-GIANT BRANCH (TRGB) DISTANCE, AND THE VALUE OF $H_{0}$}

\author{G.~A. Tammann and B. Reindl}
\affil{Department of Physics and Astronomy, University of Basel, \\
       Klingelbergstrasse 82, 4056 Basel, Switzerland}
\email{g-a.tammann@unibas.ch}

\begin{abstract}
     The light curve parameters of the normal type Ia SN\,2011fe are
     derived from the rich archive of the AAVSO. This leads, together
     with the TRGB distance modulus of $(m-M)=29.39\pm0.05$ of the
     parent galaxy M101, to maximum magnitudes of the unreddened SN
     of $M_{B} =-19.45\pm0.08$, $M_{V} =-19.46 \pm0.08$, and 
     $M_{I}=-19.25\pm0.06$ (for the standard decline rate of $\Delta
     m_{15}=1.1$). When these values are inserted into the Hubble line
     defined by 62 SNe\,Ia with $3000 < v < 20,000\kms$ --- and
     considering also four other SNe\,Ia with TRGB distances --- one
     obtains a large-scale value of the Hubble constant of
     $H_{0}=64.3\pm1.9\pm 3.2$. This value can be much improved in the
     future by using only TRGB distances of SNe\,Ia. 
\end{abstract}

\section{\MakeUppercase{%
         Introduction}}
\label{sec:1}
A single SN\,Ia with a reliable distance can be important for the
luminosity calibration of SNe\,Ia as a class and hence for the
determination of the cosmic value of $H_{0}$. The Hubble diagram of
SNe\,Ia out to $20,000\kms$ \citep[][in the following RTS\,05]{RTS:05}
shows that their intrinsic luminosity dispersion is $0.14\mag$ --- after
corrections for internal absorption, the decline rate, and if
necessary for the $z$-dependent $K$-effect. From this follows that the
luminosity of only one SN\,Ia determines the distances of all SNe\,Ia
--- and hence the value of $H_{0}$ --- to within a statistical error of
$0.14\mag$ or 7\% in linear distance.

     The SN\,Ia 2011fe occurred in the outer region of M101. Since
the Cepheid distance of this galaxy is rather ambiguous (see
Section~\ref{sec:3} below), we rely on the tip of the red-giant branch
(TRGB) as the most secure distance indicator for this galaxy. Attempts
to calibrate other SNe\,Ia with TRGBs have been made before
(\citealt{TSR:08a}; see also \citealt{Mould:Sakai:09b}). SN\,2011fe is
particularly favorable for the purpose because its light curve
parameters are well determined and its internal absorption is close to
zero. 

     The method of the TRGB seems problem-free if applied to old halo
populations. The absolute magnitude $M^{*}_{I}$ of the TRGB is well
determined through 22 galaxies with RR~Lyr star distances.
The result of $M^{*}_{I} = -4.05\pm0.02, \sigma = 0.08$
\citep{TSR:08a}, agrees exactly with two earlier, independent
determinations by \citet{Sakai:etal:04} and \citet{Rizzi:etal:07}. 
The color/ metallicity dependence of the TRGB has been discussed
several times with contradictory results. But in the relevant color
range around  $(V\!-\!I)^{*}=1.6$ the variation is not larger than
$0.05\mag$ \citep{TSR:08a} and is neglected in the following.
Also internal absorption poses a minimum problem in the halo fields.
The most severe problem of the TRGB are very red stars like young
AGBs, that can become brighter than the TRGB in dense fields of mixed
populations. In that case the TRGB may get drowned by contaminating
stars, as further discussed in Section~\ref{sec:3}. 
This always leads to an underestimate of the distance. TRGB distances
are so far restricted to $\sim\!22\;$Mpc \citep{Schweizer:etal:08},
but M101 lies well within the reach.

\section{THE SUPERNOVA 2011\lowercase{fe}}
\label{sec:2}
\begin{figure*}[t]
   \epsscale{0.75}
\plotone{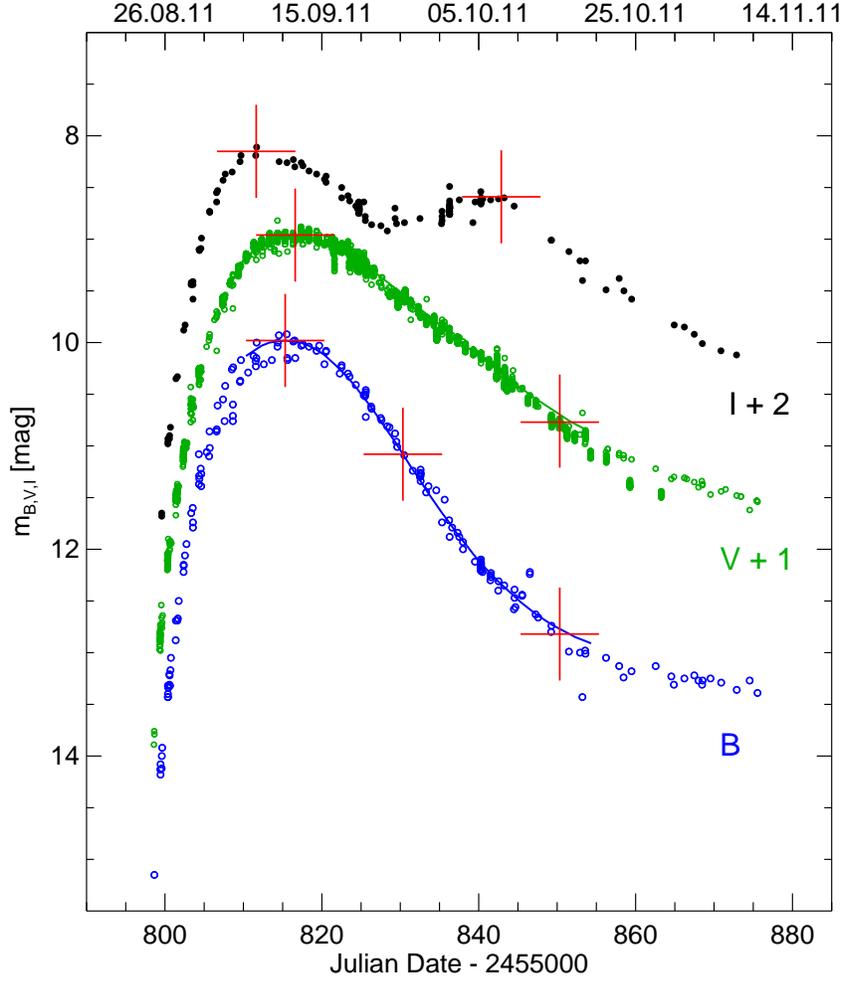}
  \caption{The light curve of SN\,Ia 2011fe in M101 in $B$, $V$, and
    $I$ from AAVSO data. The standard light curves in $B$ and $V$ of
    SNe\,Ia with $\Delta m_{15}=1.1$ are overplotted; they are taken
    from \citet{Leibundgut:88}. The maximum epochs and those 15 and 35
    days after B maximum are marked. The lightcurves in $V$ and $I$
    are shifted by $+1$ and $+2\mag$, respectively.} 
  \label{fig:01}
\end{figure*}

The SN\,2011fe was discovered in M101 by the Transient Palomar Factory
on August 24, 2011, and classified as a normal type Ia
\citep{Nugent:etal:11}. The American Association of Variable Star
Observers (AAVSO) has collected 190 $B$, 4143 $V$, and 252 $I$
magnitudes until November 10, 2011, from 48 different sources.
They establish the exceptionally well defined light curves shown in
Figure~\ref{fig:01}.  Overplotted are the well fitting standard light
curves in $B$ and $V$ for SNe\,Ia with a decline rate of $\Delta
m_{15}=1.1$ \citep{Leibundgut:88}. The relevant light curve parameters
are shown in Table~\ref{tab:01}.

\begin{deluxetable*}{llll}
\tablewidth{0pt}
\tabletypesize{\scriptsize}
\tablecaption{Properties of SN\,2011fe (corrected for Galactic
  absorption)\label{tab:01}}  
\startdata
\tableline
\tableline
\noalign{\smallskip}
Mag. at max. $=m^{\rm corr}$\tablenotemark{1)}  & $B=9.94\pm0.05$             & $V=9.93\pm0.02$                  & $I=10.13\pm0.04$                \\
Date (J.\,D. 2455000+)                          & $815.3\pm0.5$               & $816.6\pm0.5$                    & $811.6\pm0.5$                   \\
Color at max\tablenotemark{2)}                  & $(B\!-\!V)=0.01\pm0.06$     & $(B\!-\!V)_{35}=1.05\pm0.06$     & $(V\!-\!I)=-0.20\pm0.05$        \\
Reddening                                       & $E(B\!-\!V)=0.03\pm0.06$    & $E(B\!-\!V)_{35} = -0.06\pm0.06$ & $E(V\!-\!I)=0.07\pm0.05$        \\
Decline rate                                    & $\Delta m_{15}=1.10\pm0.05$ &                                  &                                 
\enddata
\tablenotetext{1)}{see text}
\tablenotetext{2)}{defined as $(B_{\max} - V_{\max})$ and $(V_{\max} -
  I_{\max})$. $(B\!-\!V)_{35}$ is the color 35 days after $B$ maximum.}
\end{deluxetable*}

     The reddenings in Table~\ref{tab:01} are the differences between
the observed colors and the intrinsic colors of SNe\,Ia with $\Delta
m_{15}=1.1$. The intrinsic colors in function of $\Delta m_{15}$ are
derived from 34 unreddened SNe\,Ia in E/S0 galaxies and outlying SNe\,Ia
in spirals (\citeauthor*{RTS:05}). The color $(B\!-\!V)$ 35 days
after $B$ maximum was also used as a fiducial reference point
following \citet{Jha:etal:07} and others. The excesses
$E(B\!-\!V)_{35}$ and $E(V\!-\!I)$ are converted into values of
$E(B\!-\!V)$ using empirical relations for SNe\,Ia from
\citeauthor*{RTS:05}. The resulting mean value is
$\langle E(B-V)\rangle=0.02\pm0.03$. This is so close to zero that it
is assumed that SN\,2011fe suffers no absorption within M101. Note
that any absorption could make the intrinsic luminosity of SN\,2011fe
only brighter. 

$V_{\max}$ occurred $1.3\pm0.7\;$days after $B_{\max}$ as compared
with a 3-day delay given by \citet{Leibundgut:88}. The SN peaked in
$I$ 3.7 days before $B_{\max}$. The secondary peak of the light curve
in $I$ at $10.59\mag$ came 31.6 days after the primary maximum. 

     Since SN\,2011fe has the fiducial decline rate of
$\Delta m_{15}=1.1$ and since it is affected by minimum internal
absorption, the corrected magnitudes $m^{\rm corr}$ are the same as
the observed magnitudes (after correction for Galactic absorption).
This makes SN\,2011fe a particularly valuable luminosity calibrator
because it minimizes errors introduced by absorption corrections and
the homogenization to $\Delta m_{15}=1.1$.

\section{\MakeUppercase{The distance of M101}}
\label{sec:3}
The TRGB of M101 has been detected in an outlying field at
$I^{*}=25.39\pm0.04$ \citep{Sakai:etal:04}. \citet{Rizzi:etal:07}
found $I^{*}=25.30\pm0.08$. (The magnitudes are corrected throughout
for Galactic absorption following \citet{Schlegel:etal:98}). The mean
value of $I^{*}=25.35\pm0.04$ gives together with the calibration from
above a true distance modulus of M101 of $(m-M)=29.39\pm0.05$, which
we adopt. 
        
     \citet{Shappee:Stanek:11} have suggested $I^{*}$ to lie at
$24.98\pm0.06$ for a reference color of $(V\!-\!I)^{*}=1.6$, yet the
magnitude is clearly too bright. The value is measured in two inner
fields of M101. The authors have taken the precaution to exclude the
stars within $4'.75$ from the center, but in spite of this the field
contains a mixture of old and young stars. Some of the latter are very
bright and red and hide the true TRGB. The authors' edge detection
function shows a weaker maximum at $I^{*}\sim25.3$ which could be the
signature of the true TRGB in agreement with \citet{Sakai:etal:04} and
\citet{Rizzi:etal:07}.

     Several similar cases are known. Already \citet{Sakai:etal:04}
have shown in case of NGC\,3621 that the TRGB is unambiguously
detected in halo fields, whereas an inner field contains so many
equally red, but brighter stars that the TRGB becomes undetectable. A
good illustration is also provided by NGC\,4038, where 
\citet{Saviane:etal:08}, considering a field containing clear tracers
of a young population, suggested a TRGB magnitude that is
$\sim\!0.9\mag$ brighter than the convincing TRGB found by
\citet{Schweizer:etal:08} in another field of this galaxy that does
not show evidence for a young population.  

     We suspect that also the claimed TRGBs of NGC\,3368 at $I^{*}=25.66$ 
\citep{Mould:Sakai:09a} and NGC\,3627 at $I^{*}= 25.77$
\citep{Mould:Sakai:09b} are determined by very red and luminous stars
of a younger population. The two galaxies are considered to be bona
fide members of the Leo~I group where four other members have a mean 
TRGB magnitude about $0.7\mag$ brighter ($\langle I^{*}\rangle =
26.42$). Also their implied TRGB distances of $(m-M)=29.71$ and 29.82
are smaller by $\sim\!0.65\mag$ than their respective Cepheid
distances of 30.34 and 30.50 \citep{Saha:etal:06}. Finally the SNe\,Ia
1998bu in NGC\,3368 and 1989B in NGC\,3627 would become fainter by
$\sim0.7\mag$ with the proposed TRGB distances than SN\,2011fe in
M101. Considering the luminosity dispersion of SNe\,Ia of only
$0.14\mag$, this corresponds to a $4.7\sigma$ discrepancy.

     For the reasons given we discard the inner-field TRGB detection
of M101 by \citet{Shappee:Stanek:11}.

     Unfortunately Cepheids cannot be used to buttress the TRGB
distance of M101, although, in general, Cepheid and TRGB moduli
agree to within $0.05\mag$ on average with a dispersion of
$\sim\!0.08\mag$ \citep{Sakai:etal:04,Rizzi:etal:07,TSR:08b}. The
problem of M101 is that the Cepheids in an outer, metal-poor field
\citep{Kelson:etal:96} and in two inner, metal-rich fields
\citep{Shappee:Stanek:11} yield discordant distances. A recent
discussion of these Cepheids gives $(m-M)=29.28\pm0.05$ for the outer
field (in acceptable agreement with the TRGB distance) and
$29.14\pm0.01$ (statistical error) for the inner fields \citep{TR:11}.
The discrepancy of $0.1-0.2\mag$ is not due to the specific choice of
the period--luminosity (P-L) relation, but is confirmed by several
authors using different P-L relations
\citep[see][their Table~3]{Sakai:etal:04}.

\section{THE CALIBRATION OF $H_{0}$}
\label{sec:4}
Sixty-two SNe\,Ia with $3000<v<20,000\kms$ define a Hubble relation of
the form 
\begin{equation}
     \log H_{0} = 0.2M^{\rm corr}_{\lambda}(\max) + C_{\lambda} + 5,
\label{eq:01}
\end{equation}
where $C_{B}=0.693\pm0.004$, $C_{V}=0.688\pm0.004$, and
$C_{I}=0.637\pm0.004$. These values are the intercepts of the
respective Hubble lines for an adopted $\Lambda$CDM model with
$\Omega_{\rm M}=0.3$ and $\Omega_{\Lambda}=0.7$ (\citeauthor*{RTS:05}, eq.~26).
  
     The corrected apparent maximum magnitudes of SN\,2011fe in
Table~\ref{tab:01} give together with the TRGB modulus of M101 from
Section~\ref{sec:3} the absolute magnitudes 
$M_{B}^{\rm corr}= -19.45\pm0.07$, 
$M_{V}^{\rm corr}= -19.46\pm0.05$, and
$M_{I}^{\rm corr}= -19.26\pm0.06$.
These values inserted into Equation~(\ref{eq:01}) give large-scale
values of
$H_{0}(B)= 63.5\pm2.2$,
$H_{0}(V)= 62.5\pm1.6$, and 
$H_{0}(I)= 61.0\pm1.8$. 
The weighted mean value is $H_{0}=62.2\pm0.9$. In view of the internal
luminosity dispersion of SNe\,Ia of $0.14\mag$ we adopt a large
statistical error of 
\begin{equation}
     H_{0}=62.2\pm4.4
\label{eq:02}
\end{equation}
from M101 and its SN\,Ia alone. The systematic error is negligible
in comparison. 

     There are four additional SNe\,Ia with direct or circumstantial
TRGB distances. They are compiled in Table~\ref{tab:02} together with
SN\,2011fe. The galaxy, its type, and its SN are listed in
Columns~1--3. The apparent maximum $V$ magnitudes are from
\citet[][SN\,1989B]{Wells:etal:94} and from Section~\ref{sec:2}
(SN\,2011fe). The remaining magnitudes are from \citet[][since only
the $B$ magnitudes are published, the $V$ magnitudes were kindly
provided by the authors]{Altavilla:etal:04}. 
The magnitudes $m_{V}^{\rm corr}$ in Column~4, corrected for
Galactic and internal absorption and normalized to $\Delta
m_{15}=1.1$, are listed in Table~2 of \citeauthor*{RTS:05}; their
errors are estimated to include also the errors of the internal 
absorption corrections and of the $\Delta m_{15}$ normalization.
The mean TRGB distances, assuming $M^{*}_{I}=-4.05$ throughout (see
Section~\ref{sec:1}), and their sources are in Columns~5 and 6. The
resulting corrected absolute magnitudes $M_{V}^{\rm corr}$ are in
Column~7. 

\begin{deluxetable*}{lllrclc}
\tablewidth{0pt}
\tabletypesize{\scriptsize}
\tablecaption{The luminosity calibration of SNe\,Ia based on TRGB
  distances\label{tab:02}}  
\tablehead{
 \colhead{Galaxy}             &  
 \colhead{Type}               & 
 \colhead{SN}                 & 
 \colhead{$m^{\rm corr}_{V}$} & 
 \colhead{$(m-M)_{\rm TRGB}$} &
 \colhead{Ref.}               &
 \colhead{$M^{\rm corr}_{V}$} \\ 
 \colhead{(1)}  & \colhead{(2)}  &
 \colhead{(3)}  & \colhead{(4)}  &
 \colhead{(5)}  & \colhead{(6)}  &
 \colhead{(7)}  
} 
\startdata
NGC\,5457  & Sc  & 2011fe & $ 9.93\pm0.08$ & $29.39\pm0.05$ & 1,2 & $-19.46\pm0.09$ \\
NGC\,5253  & Am  & 1972E  & $ 8.49\pm0.08$ & $27.79\pm0.10$ & 1,2 & $-19.30\pm0.13$ \\
IC\,4182   & Im  & 1937C  & $ 8.99\pm0.15$ & $28.21\pm0.05$ & 1,2 & $-19.22\pm0.16$ \\
NGC\,3368  & Sab & 1998bu & $11.04\pm0.10$ & $30.47\pm0.10$ & 3   & $-19.43\pm0.14$ \\
NGC\,3627  & Sb  & 1989B  & $10.95\pm0.12$ & $30.47\pm0.10$ & 3   & $-19.52\pm0.16$ \\
\tableline
\noalign{\smallskip}
\multicolumn{6}{l}{unweighted mean}             & $-19.38\pm0.05$ \\
\multicolumn{6}{l}{  weighted mean}             & $-19.40\pm0.06$ 
\enddata
%
\tablerefs{
   (1) \citealt{Sakai:etal:04},
   (2) \citealt{Rizzi:etal:07},
   (3) Galaxy assumed to be a member of Leo~I group (see text).
}                          
\end{deluxetable*}

The apparent magnitude of SN\,1937C depends heavily on the work of
\citet{Schaefer:96} who has transformed the old photometry into a
modern system. The SN is in addition a slow decliner and requires a
relatively large $\Delta m_{15}$ correction. SNe\,Ia 1989B and 1998bu
have large corrections for internal absorption ($E(B\!-\!V)\sim0.3$).
Moreover, the TRGB distances of their host galaxies, NGC\,3627 and
3368, have been rejected in Section~\ref{sec:3}. Instead it is assumed that
they are genuine members of the Leo~I group whose mean distance of
$(m-M)=30.47\pm0.06$ is secure from four other group members with good
TRGB magnitudes. These galaxies are NGC\,3351, 3377, 3379, and 3384
with TRGB determinations by 
\citet{Sakai:etal:97,Sakai:etal:04},
\citet{Rizzi:etal:07}, and
\citet{Mould:Sakai:09a}. 
The group membership is particularly compelling for NGC\,3368 (M96)
which lies between the two TRGB galaxies NGC\,3351 (M95) and
NGC\,3379 (M105). 

     The mean value of $M_{V}^{\rm corr}$ in Table~\ref{tab:02} is
remarkably stable. The dispersion of $0.13\mag$ agrees with what one
expects from distant SNe\,Ia. The weighted mean is
$\langle M_{V}^{\rm corr}\rangle = -19.40\pm0.06$,
the unweighted mean is only $0.02\mag$ fainter, as is the mean if
SN\,2011fe is omitted. If one suspects SN\,1937C because of its old,
yet modernized photometry and if one excludes SNe 1938bu and 1989B
because their TRGB distances depend on group membership, the remaining
two SN\,2011fe and SN\,1972E give $\langle M_{V}^{\rm corr}\rangle =
-19.41\pm0.07$. The gain of the overall mean over the determination
from only SN\,2011fe is that the statistical error is now much
reduced. We adopt the above weighted mean luminosity of the five
SNe\,Ia that, inserted in equation~(\ref{eq:01}), gives 
\begin{equation}
        H_{0} = 64.3\pm1.9\pm3.2.
\label{eq:03}
\end{equation}
The systematic error depends now mainly on the assumption that
NGC\,3627 and 3368 do \textit{not} lie at their suggested TRGB
distances, but that they are actual members of the Leo~I group.
For the additional reasons given in Section~\ref{sec:3} we are convinced
that the attribution is correct within the (small) depth effect of the
group. We therefore estimate the systematic error to be not more than
5\%. 

     For comparison it is noted that the TRGB distances of 78 field
galaxies with $v_{220}>280\kms$ ($v_{220}$ is the velocity corrected
for a local Virgocentric infall vector of $220\kms$) give a local
value of $H_{0}=62.9\pm1.6$ (\citeauthor*{TSR:08b}). A slightly
augmented sample of \citet{Saha:etal:06}, comprising now 29 Cepheid
distances with $280<v_{220}<1600\kms$, yields $63.4\pm1.8$
(\citeauthor*{TSR:08b}). Twenty Cepheid-calibrated SNe\,Ia with 
$v_{220}<2000\kms$ lead to a still quite local value of
$H_{0}=60.2\pm2.7$ (\citeauthor*{TSR:08b}). A large-scale value comes
from the summary paper of the \textit{HST} Supernova Project, based on
ten SNe\,Ia with Cepheid distances and a total of 62 SNe\,Ia with
$2000<v<20,000\kms$, that resulted in
$H_{0}=62.3\pm1.3\pm5$ \citep{STS:06}.
 
     Population~I and Population~II distance indicators point here
consistently to a value of $60<H_{0}<65$. However other authors have
found $H_{0}>70$ from a different --- as we believe inadequate ---
treatment of the period-color and period-luminosity relations of
Cepheids 
\citep[e.g.,][]{Freedman:etal:01,Riess:etal:09}. 
This illustrates the importance of an independent distance scale based
on Population~II objects only. 
 
     Additional support for a rather low value of $H_{0}$ comes from
\citet{Reid:etal:10} who have derived, by combining the catalog of
luminous red galaxies (LRG) from the Sloan Digital Sky Survey DR7 with
the 5-year WMAP data and the Hubble diagram of the SN\,Ia Union
Sample, a value of $H_{0}=65.6\pm2.5$ on the assumption of a 
$\Lambda$CDM model.

\section{\MakeUppercase{Discussion}}
\label{sec:5}
The importance of the RR~Lyr star-calibrated TRGB is that it uses
exclusively Population~II objects and provides an independent check on
the Population~I Cepheid distance scale. The determination of Cepheid
distances is time-consuming and complex. The metallicity dependence of
their colors makes the determination of internal absorption uncertain
and implies also that their period-luminosity relation changes with
metallicity \citep{ST:08}. Notwithstanding the general agreement,
already cited, of Cepheid and TRGB distances, the discrepancies of the
Cepheid distances of the two galaxies, where these variables were
observed in two separate fields, viz. M101 (see Section~\ref{sec:3}) and
even more pronounced in case of NGC\,4258 \citep{TR:11}, are
unexplained. Alarming are also the Cepheids in NGC\,1309 by
\citet{Riess:etal:09b} whose observed $(V\!-\!I)$ color --- uncorrected
for internal reddening --- is \textit{bluer} by $0.15\mag$ on average 
than that of any other long-period Cepheids \citep{TR:11}. Also the
Cepheids of NGC\,3021 \citep{Riess:etal:09b} are very blue. Other blue
Cepheids may be concealed by internal reddening. The properties of
blue Cepheids must be influenced by one or more hidden parameters, as
for instance the Helium content. Since the luminosity of these
Cepheids is unknown, they cannot be used for a distance determination.

     The difficulties with some Cepheids illustrate the need for an
independent confirmation of the large-scale distance scale. The test
can be provided by TRGB distances which carry the distance scale --- if
used for the luminosity calibration of SNe\,Ia --- to the limit where
SNe\,Ia can be observed. 

     The present result will be much improved by future TRGB distances
of galaxies that have produced a SNe\,Ia. It is desirable that the
search for the TRGB in NGC\,3368 and 3627 in the Leo~I group will be
repeated in some uncontaminated halo fields. The next easiest targets
at present are the Virgo cluster galaxies NGC\,4526 with SN\,1994D and
NGC\,4639 with SN\,1990N. Slightly more difficult are the Fornax
cluster galaxies NGC\,1316 with three(!) SNe\,Ia (1980N, 1981D, and
2006dd) and NGC\,1380 with SN\,1992A. These SNe, except SN\,1981D, are
blue and hence little affected by internal absorption. The Virgo and
Fornax galaxies will be rather nearer than NGC\,4038 where
\citet{Schweizer:etal:08} have demonstrated that the TRGB can be
reached. 

     It is foreseeable that the route to a precision value of $H_{0}$
through the TRGB and SNe\,Ia will become highly competitive with any
Cepheid-based distance scale.

\begin{acknowledgements}
We thank Dr. Shoko Sakai for valuable informations. We have made
extensive use of data of the AAVSO.
\end{acknowledgements}




\begin{thebibliography}{}
%
\bibitem[Altavilla et~al.(2004)]{Altavilla:etal:04}
   Altavilla, G., et~al. 2004, 
   \mnras, 349, 1344
%
\bibitem[Freedman et~al.(2001)]{Freedman:etal:01}
   Freedman, W.~L., et~al. 2001, 
   \apj, 553, 47
%
\bibitem[Jha et~al.(2007)]{Jha:etal:07}
   Jha, S., Riess, A.~G., \& Kirshner, R.~P. 2007,
   \apj, 659, 122
%
\bibitem[Kelson et~al.(1996)]{Kelson:etal:96}
   Kelson, D.~D., et~al. 1996, 
   \apj, 463, 26
%
\bibitem[Leibundgut(1988)]{Leibundgut:88}
   Leibundgut, B. 1988,
   PhD thesis, Univ. Basel
%
\bibitem[Mould \& Sakai(2009a)]{Mould:Sakai:09a}
   Mould, J., \& Sakai, S. 2009a,
   \apj, 694, 1331
%
\bibitem[Mould \& Sakai(2009b)]{Mould:Sakai:09b}
   Mould, J., \& Sakai, S. 2009b,
   \apj, 697, 996
%
\bibitem[Nugent et~al.(2011)]{Nugent:etal:11}
   Nugent, et~al. 2011,
   in press, arXiv:1110.6201
%
\bibitem[Reid et~al.(2010)]{Reid:etal:10}
   Reid, B.~A., et~al. 2010,
   \mnras, 404, 60
%
\bibitem[Reindl et~al.(2005)RTS\,05]{RTS:05}
   Reindl, B., Tammann, G.~A., Sandage, A., \& Saha, A. 2005,
   \apj, 624, 532 (RTS\,05)
%
\bibitem[Riess et~al.(2009a)]{Riess:etal:09}
   Riess, A.~G., et~al. 2009a, 
   \apj, 699, 539
%
\bibitem[Riess et~al.(2009b)]{Riess:etal:09b}
   Riess, A.~G., et~al. 2009b, 
   \apjs, 183, 109
%
\bibitem[Rizzi et~al.(2007)]{Rizzi:etal:07}
   Rizzi, L., Tully, R.~B., Makarov, D., Makarova, L., Dolphin, A.~E.,
   Sakai, S., \& Shaya, E.~J. 2007, 
   \apj, 661, 815
%
\bibitem[Saha et~al.(2006)]{Saha:etal:06}
   Saha, A., Thim, F., Tammann, G.~A., Reindl, B., \& Sandage, A. 2006,
   \apjs, 165, 108
%
\bibitem[Sakai et~al.(2004)]{Sakai:etal:04}
   Sakai, S., Ferrarese, L., Kennicutt, R.~C., \& Saha, A. 2004,
   \apj, 608, 42
%
\bibitem[Sakai et~al.(1997)]{Sakai:etal:97}
   Sakai, S., Madore, B.~F., Freedman, W.~L., Lauer, T.~R., Ajhar,
   E.~A., \& Baum, W.~A. 1997,
   \apj, 478, 49
%
\bibitem[Sandage \& Tammann(2008)]{ST:08}
   Sandage, A., \&  Tammann, G.~A. 2008,
   \apj, 686, 779
%
\bibitem[Sandage et~al.(2006)]{STS:06}
   Sandage, A., Tammann, G.~A., Saha, A., Reindl, B., Macchetto,
   F.~D., \& Panagia, N. 2006,
   \apj, 653, 843
%
\bibitem[Saviane et~al.(2008)]{Saviane:etal:08}
   Saviane, I., Momany, Y., da Costa, G.~S., Rich, R.~M., \& Hibbard, J.~E. 2008,
   \apj, 678, 179
%
\bibitem[Schaefer(1996)]{Schaefer:96}
   Schaefer, B.~E. 1996,
   \aj, 111, 1668
%
\bibitem[Schlegel et~al.(1998)]{Schlegel:etal:98}
   Schlegel, D., Finkbeiner, D., \& Davis, M. 1998,
   \apj, 500, 525
%
\bibitem[Schweizer et~al.(2008)]{Schweizer:etal:08}
   Schweizer, F., et~al. 2008, 
   \aj, 136, 1482
%
\bibitem[Shappee \& Stanek(2011)]{Shappee:Stanek:11}
   Shappee, B.~J., \& Stanek, K.~Z. 2011,
   \apj, 733, 124
%
\bibitem[Tammann \& Reindl(2011)]{TR:11}
   Tammann, G.~A., \& Reindl, B. 2011,
   in The Fundamental Cosmic Distance Scale:
   State of the Art and the Gaia Perspective, in press,
   arXiv:1112.0170
%
\bibitem[Tammann et~al.(2008a)TSR08a]{TSR:08a}
   Tammann, G.~A., Sandage, A., \& Reindl, B. 2008a,
   \apj, 679, 52 (TSR08a)
%
\bibitem[Tammann et~al.(2008b)TSR08b]{TSR:08b}
   Tammann, G.~A., Sandage, A., \& Reindl, B. 2008b,
   \aapr, 15, 289 (TSR08b)
%
\bibitem[Wells et~al.(1994)]{Wells:etal:94}
   Wells, L.~A., et~al. 1994, 
   \aj, 108, 2233
%
\end{thebibliography}
\end{document}